\documentclass[aps,amssymb,12pt,showpacs,showkeys,letterpaper]{revtex4}
\usepackage{graphicx}
\usepackage{amsmath}
\usepackage{epsfig}
\usepackage{bm}

\newcommand{\be}{\begin{equation}}
\newcommand{\ee}{\end{equation}}
\newcommand{\beq}{\begin{eqnarray}}
\newcommand{\eeq}{\end{eqnarray}}

\begin{document}
\title{Snyder Like Modified Gravity in  Newton's Spacetime }

\author{ Carlos Leiva }
 \email{cleivas@uta.cl, cleivas62@gmail.com}
\affiliation{\it Departamento de F\'{\i}sica, Facultad de Ciencias,  Universidad de
Tarapac\'{a}, Avenida General Vel\'{a}squez 1775,  Casilla 7-D, Arica, Chile}

\date{\today}

\begin{abstract}

This work is focused on searching a geodesic  interpretation of the dynamics of a particle under the effects of a
Snyder like deformation in the background of the Kepler problem. In order to accomplish that task,
a newtonian spacetime is used.
Newtonian spacetime is not a metric manifold, but allows to introduce a torsion free connection in order to interpret the
dynamic equations of the deformed Kepler problem as geodesics in a curved spacetime. These geodesics and the curvature
terms of the Riemann and Ricci tensors show a mass and a fundamental length dependence as expected, but are velocity
independent.
In this sense, the effect of introducing a deformed algebra  is examinated and the corresponding curvature terms calculated,
as well as the modifications of the integrals of motion.

\end{abstract}

\pacs{02.40.Gh, 04.60.Bc, 04.90.+e}

\keywords{Newton spacetime, Kepler problem, Snyder algebra.}

\maketitle
\section{\label{sec:Int}Introduction}

Nowadays noncommutative geometry is developing faster and faster as
a candidate to be the arena of Physics at very high energies, where
it is supposed quantum gravity should play a main role. Effectively,
it is thought that the spacetime description itself must be modified
at that scales. In that sense, main stream research on very high
energy physics and on possible candidates for a suitable quantum
gravity theory has lead to the idea of the existence of a
fundamental length. Loop Quantum Gravity, String Theory and all
their modifications and variations, propose the existence of a
minimal fundamental length that is usually identified with the
minimal size of the elements of these theories. The very existence
of such fundamental length, the same for all observers,  introduces
a new observer invariant quantity that defies the Lorentz invariance
based only on the invariance of speed of light. Hence, there are a number of
effective theories like Double Special Relativity (DSR)
\cite{Magueijo:2001cr} and Gravity's Rainbow \cite{Magueijo:2002xx},
where velocities or momenta enter in the connections and curvature terms.

A fundamental length, besides a fundamental speed, is usually
introduced through modification of the Symplectic structure of the
phase space, postulating a deformed algebra between momenta and
position operators, as in the $\kappa$-deformed algebras.

There are  different ways to introduce a minimal fundamental length,
and many of them are incompatible with Lorentz symmetry, of course
this is a  very undesirable
consequence,\cite{MirKasimov:1996hv}\cite{Djemai:2003kd}\cite{Mirza:2002wm}.

There is, however, a  way to skip the Lorentz symmetry breaking that
was  proposed by H. Snyder, \cite{Snyder}, who postulated a Lorentz
invariant modification of the Heisenberg algebra that implies
discrete spectra of the spacetime operators.

Snyder proposal researchers, being this an effective approach more than a fundamental
theory, focus their efforts on  finding ways to experimentally test
the discreteness of spacetime, introducing the modified relations of
the Snyder algebra in the quantum and in the classical realms. In
the classical version of Snyder modifications there have been some
efforts using planetary data and different approaches from Newtonian
Mechanic and General Relativity dynamics , for instance Mignemi
\cite{Mignemi:2014jya} for the last case and Romero
\cite{Romero:2003tu} and Leiva \cite{Leiva:2012az} for the former.

The classic version of Snyder algebra is based on the noncanonical
Poisson brackets:

\begin{eqnarray}
\{\bar{x}_i,\bar{x}_j\}&=& l^2L_{ij}, \\ \label{sna}
\{\bar{x}_i,p_j\}&=&\delta_{ij}+l^2p_ip_j, \nonumber \\
\{p_i,p_j\}&=&0, \nonumber
\end{eqnarray}
where $i,j=1,2,3$, and  $l$ is a parameter that measures  the deformation introduced
in the canonical Poisson brackets whit dimensions of inverse of
momentum, and $L_{ij}$ is the angular momentum matrix.

Snyder space can be realized rather successfully on constant
curvature momentum spaces (see e.g. \cite{Girelli:2010wi,{KowalskiGlikman:2002ft}}), but it's
very difficult to find spacetime versions of the Snyder space due
the non trivial dependence of the dynamic on the momentum of the
particle itself. Moreover, it's very debatable if it is possible to
implement the Snyder relations in the realm of General Relativity,
because the obtained dynamics depend in a  non linear way on the mass, so the
equivalence principle is broken and no frame it is possible to
choose in order to cancel gravity effects, not even locally.
Finally, among the difficulties that the models present it is  remarkable  the so called Soccer-Ball problem, consisting in the not clear treatment of multi-particles states that
consists on the fact that the modified Lorentz transformations  act on momenta in non linear way. (see e.g. \cite{Hossenfelder:2014ifa}).

Taken all these difficulties into account, it is possible however,
to think on implementing Snyder relations in Newton  and in
Minkowski spaces, the key is to find a good path to, at least,
represent the dynamics of a particle in a right way. It is important to mention that even General Relativity seems to be  the real and fundamental theory,
the majority of phenomena in the universe can be adequately described by Newtonian gravity like the solar system, normal stars and galaxies. It  would be then very interesting to find any clue of noncommutative Poisson brackets relations between variables in the dynamic of the (let´s say),  Newtonian Universe.

In this paper it is used a realization of the noncommutative Snyder
relations between the space an momentum variables that allows to
describe the dynamics of a particle in a Kepler central field and
describe it as geodesics in a spacetime. It is a fact that
newtonian spacetime is not a metric space, but it can be constructed
as a manifold with a connection and consisting of copies
of a spatial manifold, foliated by a parameter lineally depending on a universal time. In that way, it is
possible to find geometrical quantities that represent correctly the dynamics of
a "Snyder particle".

\section{Newton's Spacetime}

One of the main problem with Snyder and other similar proposals is that while it is common to find an interpretation of the phenomena as a curvature on the momenta space, it has not been possible to find a model of spacetime whit curvature that can explain the models. In order to attempt that, we are going to recall that it is possible to construct a Newtonian spacetime in order to include gravity as curvature. In this section the idea of a Newtonian Spacetime is constructed following the ideas presented  in \cite{Stewart}

One can't construct a metric in order to interpret the Newton gravity as curvature (see e.g.\cite { Padmabhan}). It is possible, however, to construct a quintuple $\{\mathfrak{M},\vartheta,\textsl{A},\nabla,t \}$, where $\{\mathfrak{M},\vartheta,\textsl{A}\} $ that is a smooth manifold with $\vartheta$ a standard topology and \textsl{A} an atlas. We can choose also the function $t: \mathfrak{M} \rightarrow \mathbb{R} $, and $\nabla$ satisfying:

\begin{enumerate}

\item
There is an absolute space , with $dt|_p\neq0$, everywhere for all $p \in \mathfrak{M}$. The definition of this absolute space at time $\tau$ is the set of points

\begin{equation}
S_\tau=\{p\in M / t(p)=\tau \}. \nonumber
\end{equation}

In such way that $\mathfrak{M}=\dot{\bigcup} \, S_\tau$. Where $\dot{\bigcup}$ is the disjoint union.

 \item

Absolute time flows uniformly, $\nabla \, dt=0$, everywhere.

\item

The connection $\nabla$ is torsion free.

 \end{enumerate}

So, the smooth manifold $\mathfrak{M}$ is constructed from infinite copies of the $S$ spatial set, each one labeled by the absolute time $t$.

This is not a metric space, but it's a manifold equipped  with a connection $\nabla$, usually called a Galilean Manifold. In such space it is totally possible to define a covariant derivative and Christoffel symbols, furthermore, it is possible to define the Riemann and the Ricci curvature tensors. It is not possible, however to define the Ricci scalar because one has not a metric to calculate a metric trace $g^{\mu\nu} \, R_{\mu\nu}$.

Let's write the motion equation for a probe particle in a gravity field due to another particle of mass $M$, considering, as usual,  that the movement is on a plane orbit with spatial coordinates $\rho,\theta$ and temporal coordinate $\tau$ that is a linear function of the time $t$ that label the different copies of the spatial subspace and $\dot{\tau}=1$.

\begin{eqnarray}
  \ddot{\rho}-\rho\dot{\theta}^2+\frac{GM}{{\rho}^2}\dot{\tau}^{2} &=& 0, \\
  \ddot{\theta}+\frac{2\dot{\rho}\dot{\theta}}{{\rho}} &=& 0, \\
  \ddot{\tau}=0.
\end{eqnarray}

Considering that these equations correspond to geodesics in the spacetime $\mathfrak{M}$, that are totally possible to define just with the connection, one can read the non zero Christoffel terms:

$$\Gamma^{\rho}_{\theta\theta}=-\varrho ;\hspace{1cm} \Gamma^{\rho}_{\theta\theta}=\frac{GM}{\rho^{2}} ;\hspace{1cm} \Gamma^{\theta}_{\rho\theta}=\frac{1}{\rho}.$$

The non zero  Riemann and Ricci curvature tensor components are:

$$R^{\rho}_{\tau\rho\tau}=\frac{-2GM}{\rho^3}; \hspace{1cm} R^{\theta}_{\tau\theta\tau}=\frac{GM}{\rho^3};$$

$$R_{\tau\tau}=\frac{-GM}{\rho^3}.$$

It's worth to note that the Ricci component $R{\tau\tau}$ coincides with the expression of the $-\nabla^2\phi$, where $\phi$ is the Newtonian potential. So, the Poincar\'{e} equation for Newton gravitation law can be written as:

$$R_{\tau\tau}=4 \pi G \sigma,$$

with $\sigma$ the mass density generating the gravitational potential.

\section{Kepler potential in terms of noncommutative variables}

After the proposal of Battisti and Meljanac \cite{Battisti:2010sr}, it is possible to find a a covariant realization of Snyder geometry (\ref{sna}).
Considering the noncommutative variables

\begin {equation}
\tilde{x}_\mu=x_{\mu}\varphi_{1}(A)+l^2(xp)p_{\mu}\varphi_{2}(A),
\end{equation}
where $\mu, \nu=0,1,2,3$ ,$ \varphi_{1}$ and $\varphi_{2}$ are functions of the dimensionless quantity $A=l^2p^2$. The function $\varphi_{2}$ depends on $\varphi_{1}$ by:

\begin{equation}
\varphi_{2}=\frac{1+2\dot{\varphi_{1}}\varphi_{}}{\varphi_{1}-2A\dot{\varphi_{1}}}.
\end{equation}

It is possible to set two realizations for the noncommutative Snyder geometry. One from Snyder himself if one set $\varphi_{1}=1$, that implies that $\varphi_{2}=1$, and the second one from Maggiore \cite{{Maggiore:1993rv},{Maggiore:1993zu}}, where $\varphi_{1}=\sqrt{1-sp^2}$ and $\varphi_{2}=0$.

The 3-D Kepler potential   can be written in terms of spatial noncommutative vectorial variable $\tilde{\textbf{x}}$:

\begin{equation}
\textbf{V}=-\frac{MG}{\sqrt{\tilde{\textbf{x}}^2}},
\end{equation}

  This can be implemented then, choosing either Snyder or Maggiore realization  in terms of the commutative space variables
 $x,p$ and at order  $l^2$. In both cases one has:

\begin{equation}
\textbf{V}(\textbf{x})=-\frac{\kappa}{\sqrt{\textbf{x}^2+2l^2(\textbf{x}\textbf{p})^2}}, \label{Pot}
\end{equation}

Now, for calculations we need to choose coordinates. We are dealing with a central  force even in this deformed version of the Kepler problem, so we can use a $3-D$ spatial spherical  coordinates $(\rho, \theta, \varphi)$ version. Doing so and identifying at this moment the momentum as $m\dot{\textbf{x}}$, we obtain:

\begin{eqnarray}
\textbf{x}&=&\rho \hat{\rho}, \\
\textbf{p}&=&m(\dot{\rho}\hat{\rho}+\rho
\dot{\theta}\hat{\theta}+\rho\dot{\varphi}\sin(\theta)\hat{\varphi},
\end{eqnarray}
and assuming $l^2 m^{2} \dot{\rho}^2\ll 1$, the
potential for a probe particle can be written as

\begin{equation}
V(\rho)=\frac{-k}{\rho}(1-l^2 m^{2} \dot{\rho}^2). \label{potential}
\end{equation}

With this potential we can construct, as usual, a reduced Lagrangian choosing $\theta=\pi/2$ that contains the modification due a Snyder like perturbation:

\begin{equation}
L=\frac{1}{2}m[1-\frac{2l^2km}{\rho}]\dot{\rho}^2+\frac{1}{2}m\rho^2\dot{\varphi}^2+\frac{k}{\rho}. \label{Lagrangian}
\end{equation}

The Euler-Lagrange equations are:

\begin{eqnarray}
  (m-\frac{2m^2l^2}{\rho^2})\ddot{\rho}-m\rho \dot{\varphi}^2+\frac{k}{\rho^2}+\frac{m^2l^2k}{\rho^2}\dot{\rho}^2 &=& 0, \\
  \frac{d (m \rho^2 \dot{\varphi})}{dt} &=& 0.
\end{eqnarray}

One can discard the term proportional to $l^2$ in the acceleration  in front of the mass. The presence of the mass of the particle in the equations is a characteristic feature of these models, furthermore the combination $m^2l^2$ is usually present in Snyder models.  On the other hand, in order to make a better comparison with section $I$, one can focus on a Newton's  gravity problem and identify $k=GMm$, where $M$ is the mass of the particle that creates the gravitational attraction and $G$, the Newtonian constant of gravitation. Hence the equations are:

\begin{equation}
\ddot{\rho}-\rho \dot{\varphi}^2+\frac{GM}{\rho^2}+\frac{GMm^2l^2}{\rho^2}\dot{\rho}^2 = 0, \label{Snydereqs}
\end{equation}

\begin{equation}
\ddot{\varphi}+\frac{2 \dot{\rho} \dot{\varphi}}{\rho}=0. \label{Snydereqs2}
\end{equation}

Clearly, after eq. \ref{Snydereqs}, the perturbation breaks the equivalence principle, feature that is common to this kind of proposals. We cannot do General Relativity strictly speaking, but it is possible to attempt to find a  spacetime that gives account of the dynamic and the Newtonian spacetime seems to be a suitable one.

\section{Introducing Snyder perturbation in the Newton's spacetime}

Considering equations \ref{Snydereqs} and \ref{Snydereqs2} and $\ddot{\tau}=0$, one could interpret them as geodesics in the sense of Section $II$. This can be do because they effectively represent the free falling of the particle in a gravitational field (the perturbation depends on $GM$ and if $G=0$ the perturbation disappears). Because of that, they can be read as the result of the equation  $\nabla_{\textit{v}}\textit{v}=0$, being $\textit{v}$ the tangent vector to the worldline of the particle. One can then  identify the non zero Christoffel connections:

$$\Gamma^{\rho}_{\tau\tau}= \frac{GM}{\rho^2}; \hspace{1cm} \Gamma^{\rho}_{\rho\rho}= \frac{GMm^2l^2}{\rho^2}; \hspace{1cm} \Gamma^{\rho}_{\varphi\varphi}=-\rho; \hspace{1cm} \Gamma^{\varphi}_{\rho\varphi}=\frac{1}{\rho}.$$

The non zero Riemann curvature tensor componentes are:

$$R^{\rho}_{\tau\rho\tau}= \frac{-2GM}{\rho^3}+\frac{GMm^2l^2}{\rho^4}; \hspace{1cm} R^{\rho}_{\varphi \rho \varphi}=\frac{-GMm^2l^2}{\rho};
\hspace{1cm} R^{\varphi}_{\tau \varphi \tau}=\frac{GM}{\rho^3}; $$
$$\hspace{1cm} R^{\varphi}_{\rho \varphi \rho}=\frac{GMm^2l^2}{\rho^3}.$$

And finally the non zero Ricci tensor components:

$$R_{\tau\tau}=\frac{-GM}{\rho^3}[1-\frac{m^2l^2}{\rho}];\hspace{1cm} R_{\varphi\varphi}=-\frac{GMm^2l^2}{\rho};\hspace{1cm} R_{\rho\rho}=\frac{GMm^2l^2}{\rho^3}.$$

It's possible to see now that there is curvature induced on the spatial part of the manifold $\mathfrak{M}$ and, of course, one obtain the Newtonian gravity curvature interpretation in the limit $l=0$. It is also important to see that the geodesics depend on the mass of the particle, this agrees totally with the idea that free falling is not longer mass independent in this kind of models. On the other hand, it is interesting that the metric is velocity independent, it is an interesting feature because this interpretation doesn´t inherit the momenta dependence from the momenta space geometry models (see e.g. \cite{Mignemi:2016ilu}). However, the mass dependence seems to indicate that each particle sees a different spacetime and that is against the idea of a spacetime independent of a probe  particle characteristics.  But this is the characteristic of this kind of proposals, some kind of back reaction that affects the ambient spacetime of the particles in a strong non linear way.  One can also recall the electron models in condensed matter, where there is a strong interaction between the particle characteristics and the surroundings.

The interpretation of $R_{\tau\tau}$ is now  rather difficult; it contains the non perturbed part from the Newtonian gravity, but it isn't sure that just this term gives account of  a density interpretation as it was before due the existence of $ R_{\rho\rho}$ and $R_{\varphi\varphi}$. Furthermore, one can think  that $R_{\rho \rho}$ and $R_{\varphi \varphi}$ are related to some kind of density fluxes, so the interaction between the particle with a gravitating mass and the Snyder effect can be interpreted as the effect of a density-momentum quantity on this  curved spacetime and construct a new quantity $T_{\mu\nu}$ such that:

\begin{equation}
R_{\mu\nu}=4 \pi GT_{\mu\nu}.
\end{equation}

The properties of $T_{\mu\nu}$ should be examined in more detail elsewhere, but it should be done in a relativistic version of the model otherwise it has no sense. The curvature of the spatial part of $\mathfrak{M}$ is given by $R_{\rho\rho}$ and $R_{\varphi\varphi}$. Anyway it has been found a curvature interpretation of the perturbative term of Newton's gravity due to Snyder geometry in an suitable manifold that describes the particle dynamics.

\section{Integration of Equations}

The dynamical equations are very non lineal and are very difficult to solve, but one can find  integrals of motion. First of all, due to the Lagrangian \ref{Lagrangian} is not time dependent the energy is a constant and identifying the Hamiltonian as the energy operator, one has:

\begin{equation}
H=\frac{\Pi^{2}_{\rho}}{2m(1-\frac{2kml^2}{\rho})}+\frac{\Pi^{2}_{\varphi}}{2m\rho^2}=Cte.
\end{equation}

It's worth to mention that one can have the exact Hamiltonian using the version of the potential in \ref{Pot}. The second movement integral  arises noting that because the Snyder perturbation is still a pure central force, the angular momentum is conserved as can be seen from \ref{Snydereqs2}:

\begin{equation}
m\rho^2 \dot{\varphi}=\Pi_\varphi=Cte.
\end{equation}

  There is another integral of motion that is very important in central force problems, the Laplace-Runge-Lenz \textbf{A} vector that fixes the major axis of the ellipse on the orbiting plane. For a force $f(\rho)$ acting in the radial direction it is possible to write:

\begin{equation}
\frac{d}{dt}(\textbf{P}\times \textbf{L})=-mf(\rho)\rho^{2}\frac{d\hat{\rho}}{dt},
\end{equation}

where $\vec{P}$ is the linear momentum and $\vec{L}$ the angular momentum of the particle. Considering a general Kepler like force perturbed in the way proposed in this work, $f(\rho)=\frac{-k}{\rho^2}+\frac{m^2l^2k}{\rho^2}\dot{\rho}^2 $  it is possible to write:

\begin{equation}
\frac{d\textbf{A}}{dt}=-km^2l^2\frac{\dot{\rho}^2}{\rho \dot{\varphi}}\hat{\varphi}.
\end{equation}

For the integration of this equation it is possible to use the non perturbed solutions of the Kepler problem for $\rho$, $\dot{\rho}$ and $\dot{\varphi}$, depending on the characteristics of the system, for example gravitation in the case of planet orbits or electrostatic in the case of semiclassical simple atomic problems and replacing $k$ in a suitable way. This could be the focus of a next research, specially in the atomic realm because, as have been stated (e. g. \cite{Mignemi:2014jya}), the Snyder effect is plausible to have consequences at a microscopic scale only.

\section{Final Remarks}

The focus of this work was to find an interpretation of Snyder perturbation to the Kepler problem and specifically of Newton gravity in terms of curvature of a spacetime while the phase space has been examined in many other works. In this sense, a version of geodesics interpretation and  space curvature has been found, giving the possibility to extend the research to relativistic versions. The very existence of Riemann and Ricci tensors allows to visualize how the geometry of the space and the spacetime is affected due to the introduction of a independent length scale.  In fact, Newtonian space isn't  a metric space itself, but it's a starting point to research and the logical next step is to study the relativistic version and the conditions to connect with gravity. As was said before, it's not clear how to do general relativity with the DSR theories because they brake the equivalence  principle and this could be the aim of future researches.  It's important to state that the Snyder deformation is very interesting and even there are many works focused on the experimental effects,  the properties of such modified spaces are very worthy to be examined in order to better understand it. The effects of this kind of perturbation are in fact compatible with dynamics just near the Planck scale, so it seems better to concentrate efforts in the geometrical properties that could shed some light on the basis of a new interpretation of the manifold that could be appropriate to represent  spacetimes having a minimal fundamental length scale.




\begin{thebibliography}{9}

\bibitem{Magueijo:2001cr}
  J.~Magueijo and L.~Smolin,
  ``Lorentz invariance with an invariant energy scale,''
  Phys.\ Rev.\ Lett.\  {\bf 88}, 190403 (2002)
  doi:10.1103/PhysRevLett.88.190403
  [hep-th/0112090].

  \bibitem{Magueijo:2002xx}
  J.~Magueijo and L.~Smolin,
  ``Gravity's rainbow,''
  Class.\ Quant.\ Grav.\  {\bf 21}, 1725 (2004)
  doi:10.1088/0264-9381/21/7/001
  [gr-qc/0305055].

 \bibitem{MirKasimov:1996hv}
  R.~M.~Mir-Kasimov,
  ``The Snyder space-time quantization, q deformations, and ultraviolet
  divergences,''
  Phys.\ Lett.\  B {\bf 378} (1996) 181.

\bibitem{Djemai:2003kd}
  A.~E.~F.~Djemai,
  ``On noncommutative classical mechanics,''
  Int.\ J.\ Theor.\ Phys.\  {\bf 43}, 299 (2004)
  doi:10.1023/B:IJTP.0000028864.02161.a3
  [hep-th/0309034].


\bibitem{Mirza:2002wm}
  B.~Mirza and M.~Dehghani,
  ``Noncommutative geometry and the classical orbits of particles in a central force potential,''
  Commun.\ Theor.\ Phys.\  {\bf 42}, 183 (2004)
  [hep-th/0211190].


\bibitem{Snyder}
 H. Snyder, Quantized Spacetime, Phys.Rev.71:38-41,1947

\bibitem{Mignemi:2014jya}
  S.~Mignemi and R. Strajn,
  ``Snyder dynamics in a Schwarzschild spacetime,''
  Phys.\ Rev.\ D {\bf 90}, no. 4, 044019 (2014)
  doi:10.1103/PhysRevD.90.044019
  [arXiv:1404.6396 [gr-qc]].

\bibitem{Romero:2003tu}
  J.~M.~Romero, J.~D.~Vergara,
  ``The Kepler problem and noncommutativity,''
  Mod.\ Phys.\ Lett.\  {\bf A18}, 1673-1680 (2003).
  [hep-th/0303064].

\bibitem{Leiva:2012az}
  C.~Leiva, J.~Saavedra and J.~R.~Villanueva,
  ``The Kepler problem in the Snyder space,''
  Pramana {\bf 80}, 945 (2013)
  doi:10.1007/s12043-013-0540-5
  [arXiv:1211.6785 [gr-qc]].


\bibitem{Girelli:2010wi}
  F.~Girelli and E.~R.~Livine,
  ``Scalar field theory in Snyder space-time: Alternatives,''
  JHEP {\bf 1103}, 132 (2011)
  doi:10.1007/JHEP03(2011)132
  [arXiv:1004.0621 [hep-th]].

\bibitem{KowalskiGlikman:2002ft}
  J.~Kowalski-Glikman,
  ``De sitter space as an arena for doubly special relativity,''
  Phys.\ Lett.\ B {\bf 547}, 291 (2002)
  doi:10.1016/S0370-2693(02)02762-4
  [hep-th/0207279].


\bibitem{Hossenfelder:2014ifa}
  S.~Hossenfelder,
  ``The Soccer-Ball Problem,''
  SIGMA {\bf 10} (2014) 074
  doi:10.3842/SIGMA.2014.074
  [arXiv:1403.2080 [gr-qc]].


  \bibitem{Stewart}
   John Steward,
   "Advanced general relativity"
  Cambridge Monographs on MathematicalPhysics,
  Cambridge University Press 1991
ISBN: 0-521-44946-4, pages 46-51.

\bibitem{Padmabhan}
Thanu Padmabhan, "Gravitation Foundations and Fromtiers"
Cambridge University Press 2010
ISBN: 978-0-521-88223, pages 203-204.

\bibitem{Battisti:2010sr}
  M.~V.~Battisti and S.~Meljanac,
  ``Scalar Field Theory on Non-commutative Snyder Space-Time,''
  Phys.\ Rev.\ D {\bf 82}, 024028 (2010)
  doi:10.1103/PhysRevD.82.024028
  [arXiv:1003.2108 [hep-th]].


\bibitem{Maggiore:1993rv}
  M.~Maggiore,
  ``A Generalized uncertainty principle in quantum gravity,''
  Phys.\ Lett.\ B {\bf 304} (1993) 65
  doi:10.1016/0370-2693(93)91401-8
  [hep-th/9301067].


\bibitem{Maggiore:1993zu}
  M.~Maggiore,
  ``Quantum groups, gravity and the generalized uncertainty principle,''
  Phys.\ Rev.\ D {\bf 49}, 5182 (1994)
  doi:10.1103/PhysRevD.49.5182
  [hep-th/9305163].




\bibitem{Mignemi:2016ilu}
  S.~Mignemi and A.~samsarov,
  ``Relative-locality effects in Snyder spacetime,''
  Phys.\ Lett.\ A {\bf 381}, 1655 (2017)
  doi:10.1016/j.physleta.2017.03.033
  [arXiv:1610.09692 [hep-th]].





\end{thebibliography}
\end{document}